\documentclass[prl,english,longbibliography,twocolumn,amsmath,amssymb]{revtex4-1}
\usepackage{xcolor}
\usepackage{amsmath}
\usepackage{graphicx}
\usepackage{dcolumn}
\usepackage{bm}
\usepackage[caption=false]{subfig}
\usepackage{hyperref}
\usepackage{ragged2e}

\newcommand{\asc}{{a}_{\rm s}}

\renewcommand{\vec}[1]{\mathbf{#1}}
\usepackage{orcidlink}

\usepackage{graphicx}
\usepackage{dcolumn}
\usepackage{bm}
\usepackage{amsfonts,amssymb,amsmath,mathrsfs}
\usepackage[english]{babel}
\usepackage[utf8]{inputenc}
\usepackage[T1]{fontenc}
\usepackage{xspace}
\usepackage[normalem]{ulem}
\usepackage{subfig}

\usepackage{blindtext}

\begin{document}
	
	\title{Understanding Floquet Resonances in Ultracold Gas Scattering}
	
	\author{Christoph Dauer}
	\affiliation{Physics Department and Research Center OPTIMAS, RPTU Kaiserslautern-Landau, 67663 Kaiserslautern, Germany.}
	
		\author{Axel Pelster\orcidlink{0000-0002-5215-0348}}
	\affiliation{Physics Department and Research Center OPTIMAS, RPTU Kaiserslautern-Landau, 67663 Kaiserslautern, Germany.}
	
		\author{Sebastian Eggert\orcidlink{0000-0001-5864-0447}}
	\affiliation{Physics Department and Research Center OPTIMAS, RPTU Kaiserslautern-Landau, 67663 Kaiserslautern, Germany.}

	\date{\today}
	
	\begin{abstract}
 Scattering by a short-range potential with time-periodic interaction strength is investigated with a Floquet-scattering theory.  Sharp 
resonances occur, at which the s-wave scattering length can be 
tuned to large positive and negative values. We show that the shape of these resonances is described by a simple formula, and find that both resonance position and prefactor can be altered by the driving strength. Our approach allows to identify the physical origin of the scattering resonances as Floquet bound states with positive energies, which are
dynamically created by the drive.  This insight is valuable for a detailed analysis and
uncovers a general resource for enhanced or reduced scattering in Floquet systems.
\end{abstract}	
	\maketitle

The understanding
of interaction effects in quantum systems remains a major challenge in
many fields of physics.  In this context {\it tunable} interactions
are extremely valuable for systematic studies of correlated many-body phenomena.
Accordingly, extensive research has focused
on adjustable scattering lengths in ultra-cold quantum gases, using
Feshbach resonances \cite{Inouye1998,Timmermans1999,Vuletifmmodecuteclseci1999,Chin2000,Chin2010}, near resonant light \cite{Fedichev1996,Duine2004,Theis2004,Thalhammer2005,Nicholson2015,Thomas2018}, and microwave fields \cite{Kaufman2009,Zhang2009,Tscherbul2010,Hanna2010,Papoular2010,Smith2015,Owens2016,Sykes2017,widera}.

We now examine the possibility of using time-periodic scattering lengths as an
additional tuning mechanism \cite{widera,Smith2015,Sykes2017}.  
Time-oscillating lattices and 
interactions are known as Floquet systems and have been proposed and used in a 
large variety of setups \cite{DunlapD.H.1986,Eckardt2005,
Lignier2007,Sias2008,Zenesini2009,Salger2009,Kitagawa2010,Pollack2010,Dalibard2011,Vidanovic2011,Struck2012,Hauke2012,Aidelsburger2013,Rudner2013,Wang2014,Goldman2014,Jotzu2014,Cairncross2014,Flaeschner2017,Clark2017,Mann2017,Wang2018a,Kreil2019,Nguyen2019,FedorovaCherpakova2019,Wang2020,Fazzini2021,Fedorova2021a,Wamba2021}.
  For a one-dimensional setup it is known that a Floquet impurity potential
can have a rather dramatic effect \cite{Thuberg2016, Reyes2017,Thuberg2017a}: even an infinitesimally small
driving amplitude can completely block one-dimensional transport at special resonances.
Such a quantum resonance catastrophe has great potential
for magnetic filters \cite{Thuberg2017a} and superconducting filters \cite{Huebner2021,PhysRevA.108.023307}. 
Very recently, time-oscillating interparticle potentials were experimentally analyzed using ultra-cold $^6$Li atoms \cite{widera}, realizing
the effect of a "Floquet-Feshbach resonance" with the help of an oscillating magnetic field \cite{Smith2015,Sykes2017}.   
In this work we now seek to uncover the physical origin of such resonances from general time-periodic interactions and provide closed formulas for
the effective scattering, which will be analyzed to give straight-forward analytic predictions for the locations and prefactors of the resonant behavior.
We also predict a small imaginary part corresponding to losses into higher frequency modes.

The proposed setup can be modeled by a time-periodic s-wave scattering length
\begin{equation}
        \label{eq:TimeDepScattLength}
        a(t)=\bar{a}+ a_1 \cos(\omega t)\, ,
        \end{equation}
where the amplitude is assumed to fulfill the restriction $|a_1| < |\bar a|$.
In the center-of-mass frame of reference the s-wave collisions of ultracold atoms are
described by the Hamiltonian
    \begin{equation}
	\label{eq:FRSTHarmonicallyDrivenContact}
	H(r,t)=\frac{\hbar^2}{2 \mu}\left[-\Delta+ 4 \pi a(t) \delta^3(\vec r) \frac{\partial}{\partial r} r\right],
\end{equation}
in terms of the short-range pseudo-potential formalism \cite{Huang1957,Stock2005}
along the radial coordinate $r$ with reduced mass $\mu$. 
No assumptions are made about the underlying mechanism for the oscillating 
modulation $a(t)$, but it is instructive to consider 
prototypical values corresponding 
to ultracold $^6$Li atoms in a magnetic field \cite{widera}: 
The collision energy at $T\sim 770$\,nK is about a factor of 1000 smaller than the
typical dimer energy $E_{\bar{a}}=\hbar^2/2\mu \bar{a}^2 \approx 2\pi \hbar \! \times\! 16$\,MHz at $\bar{a}=100$\,\r{A}.  In the following we use units of 
$\hbar = \hbar^2/2\mu=1$, i.e.~measuring length scales in units of $|\bar a|$ will result in 
energies and frequencies relative to $E_{\bar a}$.

\begin{figure}
		\centering
		{\includegraphics[width=0.99\linewidth]{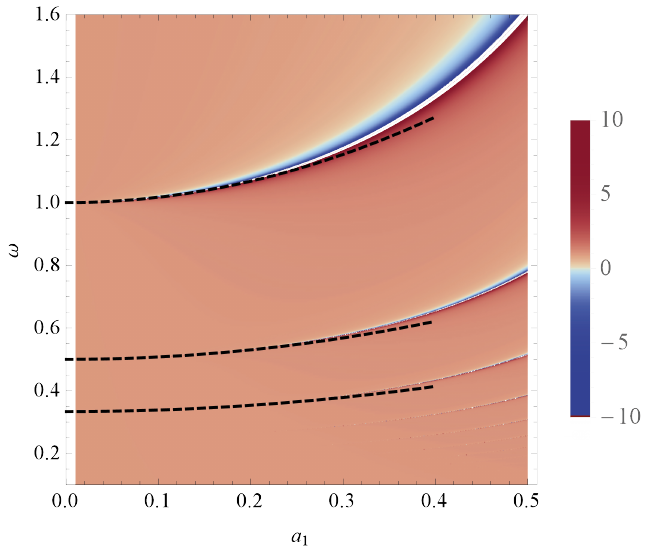}}\\~\\
		{\includegraphics[width=0.99\linewidth]{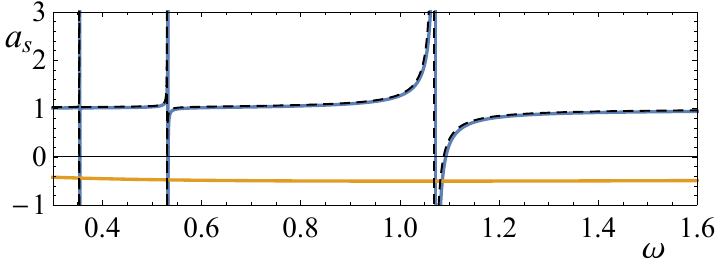}}
		
		\caption{Results for $\asc$ compared to the analytic prediction in Eqs.~(\ref{feshb2})--(\ref{Delta1}) (dashed) for $\epsilon\!\to\! 0$. Top: Real part as function of driving frequency $\omega$ and amplitude $a_1$.  Bottom: Imaginary part multiplied by 100 (orange) and real part (blue) for $a_1\!=\!0.2\,\bar a$. }
		\label{fig:longplotrealascatta101and02}
	\end{figure}

For a time periodic Hamiltonian 
$H\!=\! H_0\!+\! H_1 \cos\omega t$ the steady states 
$\psi(r,t)\!=\!e^{-i \epsilon t/\hbar} \phi(r,t)$ can be found by  
Floquet theory \cite{Shirley1965,Sambe1973,Grifoni1998,Holthaus2016,Eckardt2017},
where $\epsilon$ is the  Floquet eigenenergy.
Using the Fourier transformation $H_n\!=\! \int_0^T\! dt\,e^{i n \omega t} H(t)/T$ 
and the time-periodic Floquet mode $\phi(r,t)\!=\!\sum_n e^{-in \omega t} \phi_n(r)$, the time-dependent Schr\"odinger equation becomes an
eigenvalue equation for the Floquet quasi-energy  
$\epsilon$
\begin{equation}
        \label{eq:CoupledChannelGeneral}
        H_0 \phi_n(r)\!+\!H_1[\phi_{n-1}(r)\!+\!\phi_{n+1}(r)]/2=(\epsilon \!+\!n  \omega)\phi_n(r).
        \end{equation}
In terms of the 
pseudo-potential from Eq.~(\ref{eq:FRSTHarmonicallyDrivenContact}) 
this
yields
\begin{equation} 
(\Delta\!+\! \epsilon\!+\!n\omega) \phi_n\! =\! 4 \pi \delta^3(\vec r) \frac{\partial}{\partial r} r
(\bar a \phi_n + a_1 \frac{\phi_{n+1}\! +\!  \phi_{n-1}}{2}). 
\label{Hfloquet}
\end{equation}
We make the s-wave 
scattering ansatz
\begin{equation}
        \label{eq:AnsatzRadialWaveFunction}
\phi_n(r)= {\delta_{n,0}}\frac{\sin k_0 r}{k_0 r}+f_n  \frac{e^{i k_n r}}{r},
        \end{equation}
with an incoming plane wave $k_0$ in the $n\!=\!0$ channel. 
 Note, that for $r\!>\!0$ the usual wave equation is fulfilled
for each channel $H_0 \phi_n \!=\! k_n^2 \phi_n\! =\! (\epsilon\!+\!n\omega)  \phi_n$, 
which fixes 
the Floquet energy $\epsilon\! =\! k_0^2$ and all scattered
wave vectors $k_n \!=\! \sqrt{\epsilon\! +\!n  \omega}$.
The scattering amplitudes $f_n$  for 
$n\!>\!0$ correspond to open channels, where particles are lost, while $n\!<\!0$ are
bound states which lead to interesting resonances as we will see below.
Using $\Delta \frac{1}{r}\! =\! -4 \pi \delta^3(\vec r)$ 
we arrive at the recursion relation 
\begin{eqnarray}
\label{eq:Recursion}
k_n f_n & = & L_n \left(k_{n+1}f_{n+1}+k_{n-1} f_{n-1}\right)+{h}_n,\ \ \ \\ 
\label{eq:RecursionInhomogeniety}
{\rm where} \ \ \ \ \ \ 
L_n &  =   & -\frac{i  a_1 k_n}{2(1 + i \bar{a} k_n)}, \label{ln} \\
\ \ \ \  h_n  & = &   -i {L_n}(2 \delta_{n,0}{\bar a}/a_1 + \delta_{|n|,1}).
\end{eqnarray}
%
For a numerical analysis we could stop at this point since the linear set of 
Eq.~(\ref{eq:Recursion}) can be efficiently solved with the convergence  condition $f_{|n|\to \infty}\to 0$. 
Interestingly, as shown in 
 Fig.~\ref{fig:longplotrealascatta101and02}, large resonances 
are observed even for small $a_1$ if $\bar a >0$ which we will analyze in the following.

First, we derive a continued fraction solution  by dividing the recursion in Eq.~\eqref{eq:Recursion} for $|n|>1$ by the left-hand side, so it 
can be expressed 
in terms of fractions 
$g_n^{\pm}\!=\!k_{n}f_n/(L_{n} k_{n\mp 1}f_{n\mp 1})$ 
for positive $\!n\!>1$ and negative 
$n\!<\!-1$, respectively:
%
\begin{eqnarray}
\label{cont}
g_n^{\pm}  
& = & \frac{1}{1-{L_{n}L_{n\pm 1}g^{\pm}_{n \pm 1}}} 
= \frac{1}{ 1-\frac{L_{n} L_{n\pm 1}}{1-\frac{L_{n\pm 1 }L_{n \pm 2}}{1-\frac{L_{n\pm 2}L_{\pm n\pm 3}}{\ddots}}}}. \label{eq:ContinuedFractionnpm2}
	\end{eqnarray}
Here, we have inserted the definition of   $g_n^{\pm}$ into Eq.~(\ref{eq:Recursion}) to obtain the relation in Eq.~(\ref{cont}), which leads to the continued fraction solution 
by straight-forward iteration \cite{Brezinski1991,Simmendinger1999,Martinez2003}.
	Inserting $g_{\pm 2}^{\pm }$ from Eq.~\eqref{eq:ContinuedFractionnpm2} into 
Eq.~\eqref{eq:Recursion} for $n\!=\!0,\pm1$ the scattering amplitude is expressed in terms of an effective scattering length $\asc$ (see End Matter A)
	\begin{eqnarray}
	\label{eq:FloquetScatteringLengthCF}
	f_0& =&  -\frac{a_s}{1+i k_0 a_s} \, ,\\
{\rm where} \ \ \ \ \ \ \ \ \ \ \ \ \  
\asc & =&  \bar a+ a_1 (L_1g^+ + L_{-1}g^-_{})/2\,.\ \ \ \ \ \ \ \ \  \label{ae}
	\end{eqnarray}
Here $g^{\pm}_{}\!=\!1/(1\!-\!L_{\pm 1}L_{\pm 2}g^\pm_{\pm 2})$  are  given by the continued 
fractions for $n=\pm 1$ in Eq.~(\ref{eq:ContinuedFractionnpm2}).
The effective scattering length  $\asc$ has two 
separate contributions from the periodic driving $a_1$.    
The term $a_1 L_1 g^+$ from the open channels $n\!>\!0$ is of order $a_1^2$ and
has both real and imaginary parts. This is due to the fact that particles are
scattered into higher energy Fourier modes and are considered lost.
The term $a_1 L_{-1}g^-_{}$ from the closed channels $n\!<\!0$
is purely real since $L_{n<0} \in \mathbb{R}$ and 
shows pronounced divergences if $\bar a  \!>\!0$.
In fact, the form of the parameters $L_n$  in Eq.~(\ref{ln}) indicate 
singular points for $ -i \bar a k_n \!=\!1$. Using  $k_n\! =\! \sqrt{n \omega+\epsilon}$,
resonances will therefore occur for negative $n$ close to $n\omega+\epsilon \!\sim\! -1/\bar a^2$ if $\bar a\!  >\!0$.  Note, that for $\bar{a}\!<\!0$ no resonances are found.  
In the following, we
assume positive scattering lengths
and set $\bar a\!=\!1$, so energies and frequencies are given in units of $E_{\bar a}$. 

For the leading resonance and small $a_1$ it is sufficient to keep only the first 
fraction of $g^-_{}$ in Eqs.~(\ref{eq:ContinuedFractionnpm2}) and (\ref{ae}).  
To order $a_1^2$ the resonance is then determined by $1-L_{-1}L_{-2}=0$ to be of the standard form \cite{Chin2010}
\begin{equation}
\label{asc} 
\asc  \approx   1  +  
\frac{\Delta_1}{\omega_{1}-\omega}\, ,
\end{equation}
where in the low-energy limit $\epsilon \rightarrow 0$ position and prefactor, i.e.~width, read
$\omega_{1} =1 +  (1 + 1/\sqrt{2}\,) a_1^2$
and $\Delta_1  =  a_1^2/2$, respectively. 
In addition we get an imaginary contribution ${\rm Im} (\asc)\!\approx\!-a_1^2 \sqrt{\omega}/(4 + 4 \omega)$ from the
open channels, which is typically much smaller than the background as shown in Fig.~\ref{fig:longplotrealascatta101and02}[Bottom].  Even though higher-order terms in $a_1$ have been neglected, Eq.~(\ref{asc}) gives quite accurate results even for an amplitude of 20\% driving strength as can be seen in Fig.~\ref{fig:longplotrealascatta101and02}.
Note, that both prefactor and frequency shift of the divergence 
scale with $a_1^2$ up to higher-order corrections, which has been tested and confirmed in experiments \cite{widera}. 
The central result in Eq.~(\ref{asc}) provides a quick and accurate prediction for location 
and strength of the leading resonance, which can be tuned by both amplitude and
frequency.  
In Fig.~\ref{fig:longplotrealascatta101and02} we plot the results in the  low-energy
limit $\epsilon\!\to\! 0$
corresponding to the physical situation in ultracold gases where
$k_{\rm B} T\! \ll\! \hbar \omega$ \cite{widera}.

In order to unravel the physical origin of the resonances, we now propose an analysis in terms of Floquet bound states, which we define and predict below.
It is well known that bound states are a valuable resource for 
strongly enhanced scattering resonances in the context of 
Feshbach resonances \cite{Chin2010,Inouye1998,Timmermans1999,Chin2000,Vuletifmmodecuteclseci1999}, Fano resonances \cite{Fano1961,Miroshnichenko2010}, or bound states in the continuum \cite{Friedrich1985, Friedrich1985a}.  
The mechanism for Floquet 
resonances is more involved, since in this case the bound state itself is generated
dynamically, i.e.~the periodic drive creates the bound state, tunes
the resonance, and also provides the coupling, simultaneously.  

 Floquet bound states are defined as distinct solutions $| \phi^m \rangle$, labeled by the index $m=1,2,3, \ldots,$
 to the Schrödinger equation, where only negative Fourier modes $n\!<\!0$ are occupied. Using the Ansatz
%
\begin{equation}
\label{eq:FRSTBoundStatesModes}
\phi_{n}^{m}(r)=D_n^{m} \frac{e^{-\kappa^m_n r}}{\kappa_n^m r}
\end{equation}
with ${\kappa}_n^m=-ik_n^m=\sqrt{-\epsilon_m-n {\omega}}$ and the normalization condition
$\sum_{n<0} {|D_n^m|^2} (\kappa_n^m)^{-3}\!=\!1/(2 \pi)\! =\!{\cal N}^2$ \cite{footnote}, we obtain from 
Eq.~(\ref{Hfloquet})
\begin{eqnarray}
\label{eq:FRSTBoundStatesRecursion}
& & \hspace*{-1cm} \left(1-\kappa_n^m \right)D_n^m=\frac{{a}_1}{2}\kappa_n^m(D_{n-1}^m+D_{n+1}^m),\hspace*{0.5cm}\forall n<0 \, ,\\
\label{eq:FRSTBoundStatesBC}
& &  \hspace*{-1cm} {\rm where}\ \ \lim\limits_{n \to -\infty} D_n^m=0 \ \ 
 {\rm and} \ \  D_0^m=0.
\end{eqnarray}
Note, that Eq.~(\ref{eq:FRSTBoundStatesRecursion}) corresponds to  Eq.~(\ref{eq:Recursion}) for  $n\!<\!0$ and $D_n^m\! =\! \kappa_n^m f_n$, but with the boundary condition $\phi_0\!=\!0$.  Analogous to a static Schrödinger equation, a  set of localized solutions $| \phi^m \rangle$ can be found at discrete energies $\epsilon_m(\omega)$ with $m=1,2,3, \ldots$, depending on frequency $\omega$ and amplitude $a_1$.
Floquet bound states
exist without any
incoming wave $D^m_0\!=\!0$ due to the periodic drive.  

\begin{figure}
		\centering
		\includegraphics[width=0.99\linewidth]{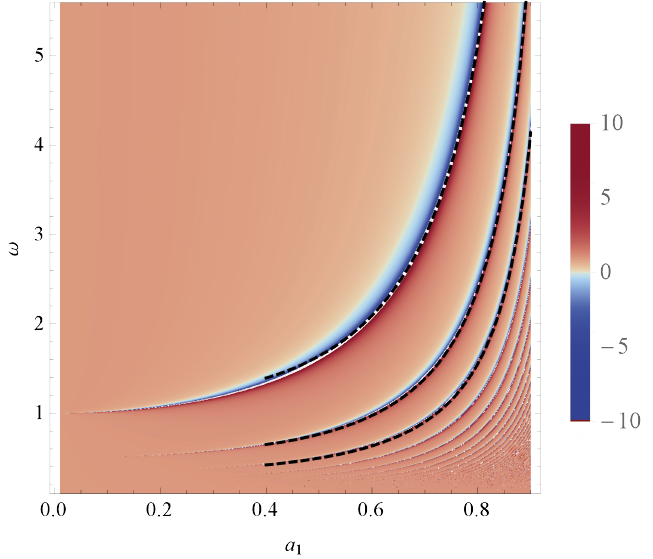}
		{\includegraphics[width=0.99\linewidth]{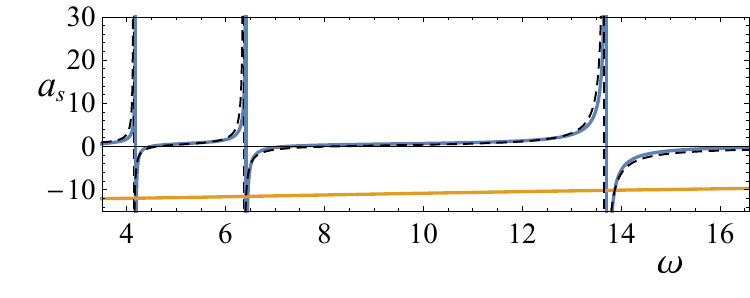}}
		\caption{Results for $\asc$ compared to the analytic prediction (black dashed) in Eqs.~(\ref{feshb2}), (\ref{eq:resonantfrequency}), and (\ref{delta}) for $\epsilon\to 0$. Top: Real part as function of frequency and amplitude.  Bottom: Imaginary part multiplied by 100 (orange) and real part (blue) for $a_1=0.9\bar a$. }
		\label{fig:omegaa1plane8ressmalla1}
	\end{figure}

Before analyzing the corresponding eigenenergies   $\epsilon_m$ in detail below, 
let us consider the implications.  The effect of bound states in the continuum on scattering resonances was discussed in a number of different undriven systems \cite{Chin2010,Inouye1998,Timmermans1999,Chin2000,Vuletifmmodecuteclseci1999,Fano1961,Miroshnichenko2010,Friedrich1985, Friedrich1985a}. In the End Matter B we now present an explicit derivation of the effect of Floquet bound states, which follows a slightly different logic from the known static cases \cite{Chin2010,Inouye1998,Timmermans1999,Chin2000,Vuletifmmodecuteclseci1999,Fano1961,Miroshnichenko2010,Friedrich1985, Friedrich1985a}.
As a result our theory accurately predicts all resonances in the effective 
scattering lengths in terms of the bound state energies $\epsilon_m(\omega)$ and the first components $D_{-1}^m$ of the normalized bound states $| \phi^m \rangle$
\begin{equation} \label{feshb}
\asc =    1-\sum_m \frac{\pi a_1^2 |D_{-1}^m|^2}{\epsilon_m-\epsilon}  +a_1L_1 g^+/2.
\end{equation}
This central result uncovers the physical origin of the observed resonances quantitatively in terms individual bound state solutions.
Note, that the locations of the resonances are given by
 $\epsilon\!=\!\epsilon_m$, which are relevant if the bound state energies are positive, i.e.~they represent bound states in the continuum \cite{Friedrich1985, Friedrich1985a}.  Expanding the frequency dependence to linear order around the incoming energy $\epsilon_m\!\approx\! \epsilon\! +\! (\omega \!-\!\omega_m)\partial_{\omega} \epsilon_m $ for each $m$ yields  the standard form of the resonances analogous to Eq.~(\ref{asc}) 
%
\begin{equation} \label{feshb2}
\asc \approx    1+\sum_m \frac{\Delta_m}{\omega_m-\omega} 
\end{equation}
with identifying $\Delta_m = \pi a_1^2 | D^m_{-1}|^2 / \partial_\omega \epsilon_m$, where the open-channel contribution $a_1L_1 g^+/2$ has been omitted.
By numerically solving the Floquet bound states, we have tested and confirmed that the predicted positions of the resonances in Eqs.~(\ref{feshb}) and (\ref{feshb2}) agree 
exactly with the continued fraction solution. We now proceed to 
find analytic approximations for $\omega_m$ and $\Delta_m$, which can be used for closed form quantitative predictions, that are 
plotted as dashed lines in Fig.~\ref{fig:longplotrealascatta101and02} and Fig.~\ref{fig:omegaa1plane8ressmalla1}.


For small $a_1$ a perturbative analysis of the bound-state solutions can be used for Eq.~(\ref{eq:FRSTBoundStatesRecursion}).
To zeroth order in $a_1$ in 
Eq.~(\ref{eq:FRSTBoundStatesRecursion})
the energies are found by $\kappa_n^m \!=\! 1$, which 
is fulfilled for discrete zeroth order energies 
$\epsilon_m\!=\! -\!1\! +\! m \omega$ and 
zeroth-order eigenstates $D_n^m\!=\!{\cal N} \delta_{m,-n}$.  
It may be surprising that bound states are
predicted in zeroth order, i.e.~for zero amplitude driving, 
but it is correct in the sense that resonances
exist in the limit of
infinitesimally small $a_1$ at frequencies $\omega_m\!=\!(1\!+\!\epsilon)/m$.
The zeroth order location of the resonances depend on the incoming energy $\epsilon$
and the order $m$, but remarkably the resonant bound wave-function has the
universal form ${\cal N}{e^{-r}}/{ r}$ from Eq.~(\ref{eq:FRSTBoundStatesModes})  
due to the underlying condition $\kappa_n^m\! =\! 1$, 
i.e.~the size is always given by the average scattering length $\bar a$.

Next, we turn to the first-order correction in $a_1$ which yields the neighboring 
coefficients $D_{\pm 1-m}^m\! =\!{\cal N} a_1/(2/\kappa^m_{\pm 1-m}\! -\!2)$.
Here we inserted the zeroth order solution $D^m_{-m}\! =\!\cal N$ into 
Eq.~(\ref{eq:FRSTBoundStatesRecursion}) for $m\!=\!-n\pm 1$ and used
$\kappa^m_{\pm 1-m}\!=\!\sqrt{1\mp \omega}$.  Finally, the second-order correction
follows from Eq.~(\ref{eq:FRSTBoundStatesRecursion}) for $m\!=\!-n$
\begin{equation}
\frac{1}{\sqrt{m\omega-\epsilon_m}}-1 \!\approx \! \frac{{a}_1^2}{4}
\left(\frac{\sqrt{1+\omega}}{1-\sqrt{1+\omega}}\!+\! \frac{\sqrt{1-\omega}}{1-\sqrt{1-\omega}}\right)
.
\end{equation}
Solving this equation for $\omega_m$ with the resonance condition of incoming energy 
$\epsilon\!=\! \epsilon_m\!=\!0$ gives 
all positions of the resonances up to order $a^2_1$
\begin{equation} \label{allres}
\omega_m \approx \frac{1}{m} + \frac{a_1^2}{2m}\left[ 2 +  \sqrt{m(m+1)}-
\sqrt{m(m-1)}\,\right]
\end{equation}
in the limit $k_0\to 0$ in agreement with recent experiments \cite{widera}. 
The leading order of the respective prefactors of the divergences as a function of $a_1$ results to \cite{DauerPhD}
\begin{equation}
\Delta_m= \frac{a_1^{2m}}{2^{2m-1}m}
\prod_{j=1}^{m-1} \left( \frac{1}{\sqrt{1-j\omega_m}}-1\right)^{-2}
\, . \label{Delta1}
\end{equation}
in good agreement with the full solution in Fig.~\ref{fig:longplotrealascatta101and02} for $a_1=0.2$.  Since $\Delta_m\propto a_1^{2m}$ becomes very small for higher orders $m\!>\!1$, we also
we need to consider stronger amplitudes beyond the perturbative approach.

Approaching larger $a_1\!\to\! \bar a$ it turns out that the 
differences between neighboring coefficients in Eq.~(\ref{eq:FRSTBoundStatesRecursion})
rapidly become smaller, so that a continuous description in terms of
differential equations can be used for the recursion relation.  The details
are presented in End Matter C with the outcome in Eq.~(\ref{eq:resonantfrequency}) that the resonance 
positions diverge
in the limit
$a_1\!\to\! \bar a$ according to
\begin{equation}
\lim_{a_1\to \bar a}\omega_m \propto \frac{1}{\sqrt{a_1 (1 - a_1)^{3}}}.
\label{divergence}
\end{equation}
The proportionality constants are given in the End Matter C as well as the corresponding expression for $\Delta_m$ in Eq.~(\ref{delta}). This results in the dashed lines in Fig.~\ref{fig:omegaa1plane8ressmalla1} in good agreement with the exact solution.
%
%

Analytical results are also possible  in the limit of large frequencies $\omega\! \gg\! \omega_{1}\! >\! E_{\bar a}$ in a regime that is beyond any resonance. 
In the recursion relation $|k_n|\!\to\! \infty$ for
$n\!\neq 0$ {in Eq.~(\ref{ln})}  leads to 
coefficients $L_n\! \to\! -a_1/(2\bar a)$ 
independent of $n$.  Therefore, also the continued fraction in Eq.~(\ref{eq:ContinuedFractionnpm2})
becomes independent of $n\neq 0$ and can be solved by a quadratic equation
$g_n^\pm \to \frac{2\bar a^2}{a_1^2}(1-\sqrt{1-a_1^2/\bar a^2}) $.  Inserting this 
into Eq.~(\ref{ae}) we have
\begin{equation}
\lim_{\omega\to \infty} \asc = \bar a \sqrt{1 -\frac{a_1^2}{\bar a^2}}. \label{avg}
\end{equation}
Hence, even for very large frequencies the driving does not average out, but reduces the
magnitude of the effective scattering length.  Note, that the result in
Eq.~(\ref{avg}) is not limited to small $a_1$ or positive $\bar a$.  It implies that
the effective interaction can be reduced by 
high frequency driving, but never outside the range
$\bar a \pm a_1$
of the time-dependent scattering in  Eq.~(\ref{eq:TimeDepScattLength}).
In the high-frequency limit, the behavior of experimental systems is typically 
stable over longer time-scales, since
there are no resonances
and the losses become negligible. 
Hence, the result in Eq.~(\ref{avg}) is suitable to be tested experimentally.

Last but not least, let us discuss the effects of higher harmonic 
terms in the time-dependence of Eq.~(\ref{eq:TimeDepScattLength}) with integer
multiples of $\omega$.  
This may be relevant  in case the scattering length is 
modified using a field in
 Feshbach resonance $a(t)/a_{\rm bg} =1-\Delta/[B(t)-B_0]$ beyond a linear regime, but 
may also offer more tuning possibilities.  The time-dependence
of the underlying field is normally under good control, so a perfect harmonic drive
can be realized, as well as large higher harmonic contributions.
Since the Hamiltonian is still time-periodic, the Floquet approach remains possible
and the recursion relation can be
modified in a straight-forward way.  Preliminary numerical results show that 
the principle structure of strong resonances is quite robust
in the presence of higher harmonics, but changes quantitatively.  This
is also true for the imaginary part, which may be larger but may also be strongly 
suppressed.  For a suppression, the phase and amplitude of the higher harmonics 
must be 
tuned so that the higher order amplitudes $f_{n>0}$ are very small, which is feasable in experiment \cite{widera}. However,
finding a general protocol to achieve  this effect was so far not possible.
We should mention at this point that an analysis of higher angular momentum 
Floquet scattering is also possible \cite{DauerPhD}, in case larger energies become 
relevant.

In summary, we have analyzed the effect of  a time-periodic scattering length
in interacting many-body systems.  The resulting effective scattering 
amplitude is highly tunable by the frequency and the amplitude of the drive.
In particular, a strong sharp enhancement is possible using even quite 
small amplitudes, which has great potential for future applications.
A closed form of the scattering amplitude was derived and analyzed, resulting
in quantitative predictions for the locations and widths of 
resonances.  Equally important, we uncovered the underlying mechanism
of the strong resonance behavior, namely bound Floquet states with positive energies, which are created dynamically by the drive.  This insight has proven to be valuable 
for obtaining analytic predictions, but also opens the door for a systematic
search, analysis, and creation of such bound Floquet states in related 
systems for enhanced tuning possibilities, changing interactions 
from small to large 
values with slight parameter changes.
\section{Acknowledgement}
We thank A. Guthmann and A. Widera for inspiring discussions.
Furthermore, we acknowledge financial support by the Deutsche Forschungsgemeinschaft (DFG, German Research Foundation) via the Collaborative Research Center SFB/TR185 (Project No. 277625399).
	\bibliography{bibpaper2}

\begin{thebibliography}{71}
\bibitem [{foo()}]{footnote}%
  \BibitemOpen
  \href@noop {} {}\bibinfo {note} {{For the scalar product time averaging and
  space integration are used, so different Fourier components $n\neq n'$ are
  orthogonal.}}\BibitemShut {Stop}%
\end{thebibliography}
\appendix
\section{End Matter A: Effective scattering amplitude } \label{app} 
In order to derive Eq.~(\ref{eq:FloquetScatteringLengthCF}) we apply
Eq.~(\ref{eq:Recursion})
for $n=\pm 1$ 
\begin{eqnarray}
k_{\pm 1}f_{\pm 1}  & =& L_{\pm 1} (k_0 f_0 + k_{\pm 2}  f_{\pm 2}-i) \nonumber \\
  & =& L_{\pm 1} (k_0 f_0 +k_{\pm 1}f_{\pm 1}L_{\pm 2}g^\pm_{\pm 2} -i) \, ,
\end{eqnarray}
where we have used 
$g_{\pm 2}^{\pm}\!=\! k_{\pm 2}f_{\pm 2}/(L_{\pm 2} k_{\pm 1}f_{\pm 1})$.  Solving this for $k_{\pm 1}f_{\pm 1}$ gives
\begin{eqnarray}
k_{\pm 1}f_{\pm 1}& =& L_{\pm 1}(k_{0}f_{0}-i)/(1-L_{\pm 1}L_{\pm 2}g^\pm_{\pm 2}) \nonumber\\
& =& L_{\pm 1}(k_{0}f_{0}-i)g^\pm_{},
\label{D1}
\end{eqnarray}
where $g^{\pm}_{}\!=\!1/(1\!-\!L_{\pm 1}L_{\pm 2}g^\pm_{\pm 2})$ are  given by the continued 
fraction in Eq.~(\ref{eq:ContinuedFractionnpm2}) for $n\!=\!\pm 1$.
For $n=0$ it follows from Eqs.~(\ref{eq:Recursion})  and (\ref{D1}) that
\begin{eqnarray}
   k_0f_0& = &  {L_0}\left(k_1f_1+k_{-1} f_{-1}-2 i \bar a/a_1\right)\label{n-1} \nonumber \\
   & = &  {L_0}\left[(k_0 f_0-i)(L_1g^+ + L_{-1}g^-_{}) -2 i \bar a/a_1\right] \nonumber\\
   & = &  {2 L_0}\left((k_0 f_0-i)(a_s-\bar a) - i \bar a\right)/a_1 \nonumber \\ 
&  =&  \frac{ -k_0}{1+ik_0\bar a}\left[ik_0f_0(\asc-\bar a)+\asc\right] \, ,
\label{d0}
\end{eqnarray}
where we have used Eq.~(\ref{ln}) and  defined 
	\begin{equation}
\asc- \bar a = a_1 (L_1g^+ +  L_{-1}g^-_{})/2. 
	\end{equation}
as in Eq.~(\ref{ae}).
Finally, solving Eq.~(\ref{d0}) for $f_0$ gives 
Eq.~(\ref{eq:FloquetScatteringLengthCF}).
\section{End Matter B: Floquet Feshbach Resonance}
The goal of this section is to evaluate the scattering contribution $a_1 L_{-1} g^-_{}$ from the closed channels in terms of the bound-state solutions $|\phi^m\rangle$ and $\epsilon_m$ as defined in Eqs.~(\ref{eq:FRSTBoundStatesModes})--(\ref{eq:FRSTBoundStatesBC}).  The calculation leading to Eq.~(\ref{ae}) defines the effective scattering length $a_{\rm s}$, but we now use the ansatz that
the bound Fourier modes $\phi_{n<0}$ in Eq.~(\ref{eq:AnsatzRadialWaveFunction})  may be written as a superposition of orthonormal bound-state solutions $|\phi^m\rangle$ in Eq.~(\ref{eq:FRSTBoundStatesModes})
\begin{equation}
    f_{n}\frac{e^{i k_n r}}{r} =\sum_m A_m \phi^m_n(r)\ \ \ \ {\rm for} \ \ n\!<\!0 . \label{ansatz}
\end{equation}
In order to determine the expansion coefficients $A_m$ we insert this ansatz into Eq.~(\ref{Hfloquet}) and use the recursion relation in Eq.~(\ref{eq:FRSTBoundStatesRecursion}) to obtain
\begin{equation}
    \sum_m (\epsilon-\epsilon_m)A_m \phi_n^m = \left[\overleftrightarrow{ \frac{\partial}{\partial r}{r} }\phi_0 \right] 2\pi a_1 \delta^3(\vec{r})\delta_{n,-1}\,.
    \end{equation}
Here the double arrow indicates an hermitian operator acting both to the right and to the left \cite{DauerPhD}, which becomes important when taking scalar products after multiplying both sides with $\phi_n^{m'*}(r)$.  
    After integrating both sides \cite{footnote}, the orthonormality of solutions $\sum_{n<0}\int d^3\vec{r} \phi_n^{m'*} (r) \phi_n^{m}(r) =\delta_{m',m}$ gives
    \begin{equation}
        A_{m'}=\frac{2\pi a_1 D^{m'*}_{-1} }{\epsilon_{m'}-\epsilon}(1+ik_0f_0).
    \end{equation}
    Finally, the relation $ k_{-1}f_{-1}=i  \sum_m A_m D_{-1}^m$ from taking $r\to 0$ in  Eq.~(\ref{ansatz}) can be inserted into Eq.~(\ref{D1}), so that
    \begin{equation}
        L_{-1} g^-_{} = -\sum_m  \frac{2\pi a_1 |D^{m}_{-1}|^2 }{\epsilon_{m}-\epsilon}
    \end{equation}
    Following the analogous steps to Eq.~(\ref{ae}) yields
 the important result Eq.~(\ref{feshb}) of the main text.

	\section{End Matter C: Continuum limit}
A continuum theory \cite{Sykes2017}
can be defined in terms of a variable $x\!=\!-n/A\geq 0$,
where we have rescaled the index according to
	\begin{equation}
	A = \left(\frac{a_1^2 \omega}{4}\right)^{1/3}\, , \ \ \ \mathcal{D}\left(x\!=\!-\frac{n}{A}\right)=(-1)^n {D_{n}}.
	\end{equation}
Using $D_{n+1} -2 D_n + D_{n-1} \approx \partial_x^2 \mathcal{D}(x)/A^2$
the bound state recursion relation in Eq.~(\ref{eq:FRSTBoundStatesRecursion})
becomes  a differential equation 
\begin{eqnarray} 
\label{eq:HighwFieldStartingEq}
  \left(-\frac{1}{\sqrt{x-\tilde\epsilon_m}}-\frac{\partial^2}{\partial^2 x}\right)
\mathcal{D}(x)& = & - E \mathcal{D}(x), \\
  {\rm where } \ \ \ \ \ \ \ \ \ \ \ \ \ \ \  \ \ \ \ 
E& =&  2 A^2 \frac{1-a_1}{a_1}.  \label{scaling}
\end{eqnarray}
Here we have used
$a_1 \kappa^m_n/(2 A^2) \!=\! \sqrt{x-\tilde\epsilon_m}$
 and defined $\tilde \epsilon_m\!=\!\epsilon_m/(\omega A)$. 
The rescaled Eq.~(\ref{eq:HighwFieldStartingEq}) has only solutions for 
discrete dimensionless eigenvalues \cite{Ishkhanyan2015}, which are given 
by $E_m\! =~\! 0.4380, 0.2632, 0.1976, 0.1617, 0.1386$ for the first five 
numerical solutions of $\epsilon_m\!=\!0$.  Inserting 
those values into Eq.~(\ref{scaling}) and solving for $\omega$
we obtain the corresponding conditions for the resonant frequency for
each solution
	\begin{equation}
	\label{eq:resonantfrequency}
	{\omega}_{m}=\frac{4}{a_1^2}\left(\frac{E_m}{2}\frac{a_1}{1-a_1}\right)^{3/2},
	\end{equation}
which shows a divergence as $a_1\!\to\! \bar a$.  Hence, the rescaling $A$ also becomes
quite large in that limit, which in turn justifies the continuous ansatz.  The corresponding prefactor the continuum limit yields \cite{DauerPhD}
\begin{equation} \label{delta}
\Delta_m = \frac{F_m}{1-a_1}, 
\end{equation}
where the first three coefficients are given by $F_m\!=\!0.212, 0.0628, 0.0313$.

\end{document}